\documentclass{ifacconf}

\usepackage{graphicx}
\usepackage{natbib}
\usepackage{amsmath,amssymb,amsfonts}
\usepackage{multirow}
\usepackage{xcolor}

\newtheorem{problem}{Problem Statement} 

\newcommand{\R}{\mathbb{R}}
\newcommand{\dhdx}{\nabla h}
\newcommand{\B}{\mathcal{B}}
\newcommand{\Kinf}{\mathcal{K}_{\infty}}
\newcommand{\Keinf}{\mathcal{K}^{\rm e}_{\infty}}

\begin{document}
\begin{frontmatter}

\title{Input-to-State Safety with Input Delay \\ in Longitudinal Vehicle Control} 

\thanks[footnoteinfo]{This research is supported by the National Science Foundation (CPS Award \#1932091), Aerovironment and Dow (\#227027AT).
The research reported in this paper and carried out at BME has been supported by the NRDI Fund (TKP2020 NC, Grant No. BME-NCS) based on the charter of bolster issued by the NRDI Office under the auspices of the Ministry for Innovation and Technology.}

\author[Caltech]{Tamas G. Molnar},
\author[UM_ME]{Anil Alan},
\author[BME]{Adam K. Kiss},
\author[Caltech]{Aaron D. Ames}, and
\author[UM_ME,UM_Civil]{Gábor Orosz} 

\address[Caltech]{
Department of Mechanical and Civil Engineering,
California Institute of Technology, Pasadena, CA 91125, USA (tmolnar@caltech.edu, ames@caltech.edu).}
\address[UM_ME]{Department of Mechanical Engineering, University of Michigan, Ann Arbor, MI 48109, USA (anilalan@umich.edu, orosz@umich.edu).}
\address[UM_Civil]{Department of Civil and Environmental Engineering, University of Michigan, Ann Arbor, MI 48109, USA.}
\address[BME]{
MTA-BME Lend{\"{u}}let Machine Tool Vibration Research Group,
Department of Applied Mechanics,
Budapest University of Technology and Economics, Budapest 1111, Hungary (kiss\_a@mm.bme.hu).}

\begin{abstract}                
Safe longitudinal control is discussed for a connected automated truck traveling behind a preceding connected vehicle.
A controller is proposed based on control barrier function theory and predictor feedback for provably safe, collision-free behavior by taking into account the significant response time of the truck as input delay and the uncertainty of its dynamical model as input disturbance.
The benefits of the proposed controller compared to control designs that neglect the delay or treat the delay as disturbance are shown by numerical simulations.
\vspace{-12pt}
\end{abstract}


\end{frontmatter}


\section{Introduction}

Control systems are often subject to strict safety requirements that must be met before deployment in practice.
As such, safety is of primary importance in vehicle control applications, where collisions must be avoided~\citep{Vasudevan2012}.
{\em Control barrier functions} (CBFs)~\citep{AmesXuGriTab2017} provide provable means to achieve safety, similarly to how Lyapunov theory yields stability.
Remarkably, one of the first applications of CBFs was adaptive cruise control~\citep{AmesCDC2014}.
Still, challenges continue to arise in safe vehicle control due to the interplay of significant time delays (the response time of vehicles, their drivers and controllers), complicated nonlinear dynamics, disturbances, and dynamically changing environments.

\begin{figure}
\begin{center}
\includegraphics[width=8.4cm]{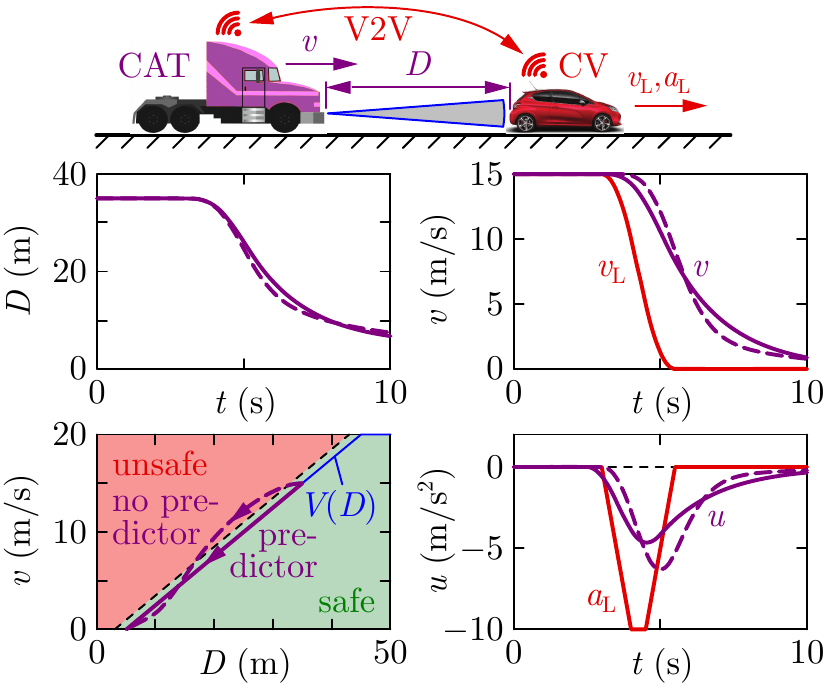}
\caption{
Safety-critical control of a connected automated truck (CAT) with input delay behind a connected vehicle (CV) during emergency braking.
The proposed predictor feedback controller maintains safety, while a baseline controller without predictor fails to do so.
} 
\label{fig:delay}
\end{center}
\end{figure}

In this work, we address safe vehicle control in the presence of both input delay and input disturbance.
Recently, safety-critical control with input delay has been addressed in discrete time \citep{necmiye2020, singletary2020control} and in continuous time for linear systems \citep{Jankovic2018, abel2019constrained} and nonlinear systems \citep{Molnar2021tcst}, including multiple and time-varying input delays \citep{abel2020constrained, Abel2021}.
Furthermore, the safety of delay-free systems with input disturbance have been investigated by robust control barrier functions \citep{jankovic2018robust} and using the notion of input-to-state safety \citep{ames2019issf, Alan2022}.
While real-life control systems may involve both delays and disturbances, their combined effect has not yet been addressed in safety-critical control --- only \cite{Seiler2022} approached this problem by treating the delay as part of the disturbance.

The contribution of our work is twofold.
First, we propose a safety-critical control method using CBFs that handles both input delay and input disturbance.
We achieve this by integrating the approaches of \cite{Molnar2021tcst} and \cite{Alan2022} that separately address delay and disturbance by relying on the concepts of predictor feedback \citep{michiels2007stability, BekKrs2013, Karafyllis2017} and input-to-state safety~\citep{ames2019issf}, respectively.
Second, we apply the proposed method to safely control a connected automated truck in the presence of a significant response time (input delay) and uncertainties in its dynamics (input disturbance).
Simulations show that the proposed approach outperforms controllers designed by neglecting the delay or treating the delay as disturbance.


\section{Motivation: Safety-critical Control of a Connected Automated Truck}

Consider the scenario in Fig.~\ref{fig:delay} where a connected automated truck (CAT) is controlled to follow a connected vehicle (CV) that shares information via vehicle-to-vehicle (V2V) communication.
Our goal is to design a controller that allows the CAT to match its speed $v$ to the speed $v_{\rm L}$ of the lead vehicle while maintaining a safe distance $D$.

To design the controller, we consider the dynamical model:
\begin{align}
\begin{split}   
\dot{D}(t) & = v_{\rm L}(t) - v(t), \\
\dot{v}(t) & = u(t-\tau) + d(t), \\
\dot{v}_{\rm L}(t) & = a_{\rm L}(t), \\
\end{split}
\label{eq:truck_dyn}
\end{align}
where $u(t)$ is the control input (commanded acceleration) of the CAT and $a_{\rm L}(t)$ is the lead vehicle's acceleration at time $t$.
Since heavy-duty vehicles have large response time due to their size and inertia, we include the input delay (powertrain delay) $\tau$ in~(\ref{eq:truck_dyn}).
Furthermore, the high-fidelity dynamics of the CAT consist of several nontrivial elements (engine and powertrain behavior, gear schedule, resistance forces, tire dynamics, etc.) that are missing from the model above.
Thus, we capture model mismatches and real-life uncertainties via the disturbance $d(t)$ in~(\ref{eq:truck_dyn}).

Our goal is to maintain safety for the CAT despite the presence of the significant delay $\tau$ and disturbance $d(t)$ in its dynamics.
Specifically, we wish to keep the distance $D$ over a minimal value
${D_{\rm sf} + T v}$,
that depends on the speed $v$ and is given by a minimum standstill distance $D_{\rm sf}$ and time headway $T$, i.e.,
${h(D(t),v(t)) := D(t) - D_{\rm sf} - T v(t) \geq 0}$, ${\forall t \geq 0}$.
We state this problem more generally below.

\begin{problem}
\textit{
For control systems of the form:
\begin{equation}
\dot{x}(t)=f(t,x(t)) + g(t,x(t)) (u(t-\tau)+d(t)),
\end{equation}
design a controller that achieves safety in that ${h(x(t)) \geq 0}$, ${\forall t \geq 0}$ holds along the solutions of the corresponding closed-loop system for a given safety certificate $h$.
}
\end{problem}

We highlight that for the CAT these quantities read: 
\begin{equation}
x =
\begin{bmatrix}
D \\
v \\
v_{\rm L}
\end{bmatrix}, \quad
f(t,x) =
\begin{bmatrix}
v_{\rm L} - v \\
0 \\
a_{\rm L}(t)
\end{bmatrix}, \quad
g(t,x) =
\begin{bmatrix}
0 \\
1 \\
0
\end{bmatrix},
\label{eq:truck_dynamics}
\end{equation}
and
\begin{equation}
h(x) = D - D_{\rm sf} - T v.
\label{eq:truck_CBF}
\end{equation}


\section{Safety with Input Delay}

The problem statement contains two main challenges for safety-critical control: input delay and input disturbance.
This section is dedicated to addressing the delay without disturbance, and disturbance is added in the next section.

\subsection{Delay-free Scenario}

First, we revisit safety-critical control for delay-free systems.
Specifically, consider a nonlinear control-affine system with state ${x \in \R^n}$, input ${u \in \R^m}$ and dynamics:
\begin{equation}
\dot{x}=f(t,x) + g(t,x) u,
\label{eq:system}
\end{equation}
where ${f: \R \times \R^n \to \R^n}$ and ${g: \R \times \R^n \to \R^{m \times n}}$ are locally Lipschitz continuous.
We consider Lipschitz continuous controllers ${k: \R \times \R^n \to \R^m}$, ${u = k(t,x)}$, and we assume that given an initial condition ${x(0) \in \R^n}$ the solution $x(t)$ of the closed-loop system exists for all ${t \geq 0}$.

To guarantee safe behavior, we define the {\em safe set} $S \subset \R^n$  that represents the set of states in which the system is considered safe.
We seek to design controller $k$ such that it enforces safety by rendering $S$ forward invariant, meaning that ${x(0) \in S \implies x(t) \in S}$, ${\forall t \geq 0}$ holds for the closed-loop system.
Specifically, we consider safe sets of the form:
\begin{equation}
S=\{x \in \R^n: h(x) \geq 0 \},
\label{eq:safeset}
\end{equation}
where ${h: \R^n \to \R}$ denotes a continuously differentiable function throughout this paper.

We endow the system with safety guarantees using control barrier functions (CBFs) that provide a practical method to synthesize  safety-critical controllers \citep{AmesXuGriTab2017}.
The main idea behind CBFs is lower-bounding the derivative of $h$ w.r.t. time, which keeps $h(x(t))$ nonnegative (i.e., ${x(t) \in S}$).
This derivative along the system~(\ref{eq:system}) reads:
\begin{align}
\begin{split}
\dot{h}(t,x,u) & = L_f h(t,x) + L_g h(t,x) u, \\
L_f h(t,x) & = \dhdx(x) f(t,x), \\
L_g h(t,x) & = \dhdx(x) g(t,x),
\end{split}
\label{eq:hdot}
\end{align}
where $L_f h$ and $L_g h$ denote the Lie derivatives of $h$ along $f$ and $g$.
With these preliminaries, we formally define CBFs.

\begin{defn}[\cite{AmesXuGriTab2017}] \label{def:CBF}
Function $h : \R^n \to \mathbb{R}$ is a {\em control barrier function (CBF)} for~(\ref{eq:system}) if there exists ${\alpha \in \Keinf}$\footnote{Function $\alpha: \R_{\geq 0} \to \R$ is of class-$\Kinf$ (${\alpha \in \Kinf}$) if it is continuous, strictly monotonically increasing, ${\alpha(0)=0}$ and ${\lim_{r \to \infty} \alpha(r) = \infty}$. Function $\alpha: \R \to \R$ is of extended class-$\Kinf$ (${\alpha \in \Keinf}$) if it is of class-$\Kinf$ and ${\lim_{r \to -\infty} \alpha(r) = -\infty}$.} such that ${\forall t \geq 0}$ and ${\forall x \in S}$:
\begin{equation}
\sup_{u \in \R^m} \dot{h}(t,x,u) > - \alpha(h(x)).
\label{eq:CBF_condition}
\end{equation}
\end{defn}

The main result from \cite{AmesXuGriTab2017} specifies a condition on the control input $u$ that provides formal safety guarantees, as summarized by the following theorem.
\begin{thm}[\cite{AmesXuGriTab2017}] \label{thm:safety}
If $h$ is a CBF for~(\ref{eq:system}), then any locally Lipschitz continuous controller $k : \R \times \R^n \to \R^m$ satisfying:
\begin{equation}
\dot{h}(t,x,k(t,x)) \geq - \alpha(h(x)),
\label{eq:safety_condition}
\end{equation}
${\forall t \geq 0}$, ${\forall x \in S}$ renders $S$ forward invariant (safe) such that ${x(0) \in S \implies x(t) \in S}$, ${\forall t \geq 0}$.
\end{thm}

The proof can be constructed based on \cite{AmesXuGriTab2017} that considers $f$ and $g$ independent of $t$.
The same proof applies here, since it is based on the theorem of~\cite{nagumo1942lage} that includes time-variant systems.
Per Theorem~\ref{thm:safety}, controllers satisfying~(\ref{eq:safety_condition}) yield safe behavior, while Definition~\ref{def:CBF} ensures that such controllers exist.
Thus,~(\ref{eq:safety_condition}) can be used as condition to synthesize safety-critical controllers.
This is often achieved by incorporating~(\ref{eq:safety_condition}) as constraint into optimization problems~\citep{AmesXuGriTab2017}.

\subsection{Control with Input Delay}

\begin{figure}
\begin{center}
\includegraphics[width=8.4cm]{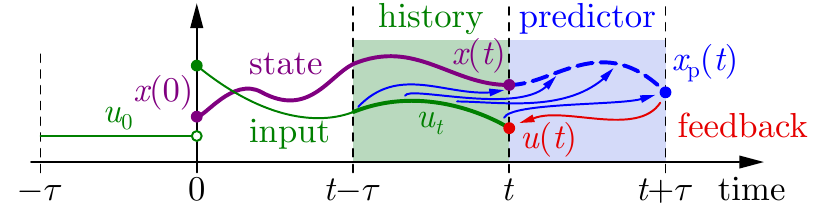}
\vspace{-6pt}
\caption{
Illustration of predictor feedback.
The state $x(t)$ (purple) and input history $u_t$ (green) are used to calculate the predicted state $x_{\rm p}(t)$ (blue) that is used as feedback to synthesize the control input $u(t)$ (red).
} 
\label{fig:predictor}
\end{center}
\end{figure}

Now we revisit the notion of CBFs with input delay, that was introduced by \cite{Molnar2021tcst} for time-invariant systems via the idea of predictor feedback \citep[ch.~15]{BekKrs2013, Karafyllis2017, michiels2007stability}.
We extend these results to the time-variant system~(\ref{eq:system}) with input delay ${\tau>0}$:
\begin{equation}
\dot{x}(t)=f(t,x(t)) + g(t,x(t))u(t-\tau).
\label{eq:system_delay}
\end{equation}
To overcome the delay in ${u(t-\tau)}$ we seek to use a prediction of the future state ${x(t+\tau)}$ as feedback.
To formalize this process, first, we characterize the solution of~(\ref{eq:system_delay}).
We begin by noting that the solution over $[t,t+\tau]$ evolves according to the history of the control input over $[t-\tau,t)$; see the illustration in Fig.~\ref{fig:predictor}.
We formally define the input history by the function ${u_t: [-\tau,0) \to \R^m}$:
\begin{equation}
u_t(\theta) = u(t+\theta), \quad \theta \in [-\tau,0),
\end{equation}
which is assumed to be bounded and continuous almost everywhere.
We shortly denote the space of such functions with $\B$.
As shown below, predictor feedback leads to controllers that depend on the input history ${u_t \in \B}$, in the form ${k: \R \times \R^n \times \B \to \R^m}$, ${u = k(t,x,u_t)}$.
Throughout the paper, we assume that given an initial condition ${x(0) \in \R^n}$ and initial input history ${u_0 \in \B}$, the solution $x(t)$ of the closed-loop system exists for all ${t \geq 0}$.

For given ${u_t \in \B}$, system~(\ref{eq:system_delay}) can be considered as a forced ordinary differential equation (ODE) over $[t,t+\tau]$.
The solution of the ODE over $[t,t+\tau]$ can be written as:
\begin{equation}
x(t+\vartheta) = \Psi(\vartheta,t,x(t),u_t), \quad \vartheta \in [0,\tau].
\label{eq:solution}
\end{equation}
That is, the solution depends on time $t$, the current state $x(t)$, and the input history $u_t$; see~Fig.~\ref{fig:predictor}.
Formally, it is given by the semi-flow
${\Psi: [0,\tau] \times \R \times \R^n \times \B \to \R^n}$:
\begin{multline}
\Psi(\vartheta,t,x,u_t)
= x + \int_{0}^{\vartheta} \Big(
f \big( t+s, \Psi(s,t,x,u_t) \big) \\
+ g \big( t+s, \Psi(s,t,x,u_t) \big) u_t(s-\tau) \Big) {\rm d}s,
\label{eq:semiflow}
\end{multline}
which can be calculated by numerical ODE solvers in practice.
Specifically, predictor feedback relies on the predicted state $x_{\rm p}(t)$ at prediction time $t_{\rm p}$:
\begin{equation}
t_{\rm p} \triangleq t+\tau, \quad
x_{\rm p}(t) \triangleq x(t+\tau) = \Psi(\tau,t,x(t),u_t).
\label{eq:predictedstate}
\end{equation}
Note that the predicted state depends on ${f(t+s,.)}$ and ${g(t+s,.)}$, i.e., on the future expressions of the dynamics in~(\ref{eq:semiflow}).
If such information is unknown, the predicted state needs to be approximated.
This leads to prediction errors, whose effect is treated as disturbance in the next section.
The predicted state is used in the following definition of CBFs for input delay systems from \cite{Molnar2021tcst}.

\begin{defn}[\cite{Molnar2021tcst}]
Function ${h : \R^n \to \mathbb{R}}$ is a {\em control barrier function (CBF)} for~(\ref{eq:system_delay}) if there exists ${\alpha \in \Keinf}$ such that ${\forall t \geq 0}$, ${\forall x \in S}$ and ${\forall u_t \in \B}$:
\begin{equation}
\sup_{u \in \R^m} \dot{h}(t_{\rm p},x_{\rm p},u) > - \alpha(h(x_{\rm p})),
\label{eq:CBF_condition_delay}
\end{equation}
where ${t_{\rm p} = t + \tau}$, ${x_{\rm p}=\Psi(\tau,t,x,u_t)}$ with $\Psi$ given by~(\ref{eq:semiflow}).
\end{defn}

This definition leads to the main result in \cite{Molnar2021tcst} that endows controllers with safety guarantees, as summarized by the following theorem.

\begin{thm}[\cite{Molnar2021tcst}] \label{thm:safety_delay}
If $h$ is a CBF for~(\ref{eq:system_delay}), then any locally Lipschitz continuous controller \linebreak${k : \R \times \R^n \times \B \to \R^m}$ satisfying:
\begin{equation}
\dot{h}(t_{\rm p},x_{\rm p},k(t,x,u_t)) \geq - \alpha(h(x_{\rm p})),
\label{eq:safety_condition_delay}
\end{equation}
with ${t_{\rm p} = t + \tau}$, ${x_{\rm p}=\Psi(\tau,t,x,u_t)}$, ${\forall t \geq 0}$, ${\forall x \in S}$ and ${\forall u_t \in \mathcal{B}}$ renders $S$ forward invariant such that ${x(t) \in S}$, ${\forall t \in [0,\tau]}$ ${\implies x(t) \in S}$, ${\forall t \geq 0}$.
\end{thm}

The proof can be constructed based on \cite{Molnar2021tcst}.
Note that safety is conditioned on ${x(t) \in S}$, ${\forall t \in [0,\tau]}$ instead of ${x(0) \in S}$, as the solution over $[0,\tau]$ is determined by the initial input history $u_0$; cf.~(\ref{eq:solution}).
If ${\tau = 0}$, these two conditions coincide and Theorem~\ref{thm:safety_delay} reduces to Theorem~\ref{thm:safety}, since ${t_{\rm p} = t}$, ${x_{\rm p} = x}$.

\vspace{-3pt}
\subsection{Application to Vehicle Control}
\vspace{-3pt}

Now we apply the above theory to design a safe longitudinal controller for the connected automated truck (CAT) in Fig.~\ref{fig:delay}.
We describe the dynamics of the CAT by~(\ref{eq:truck_dynamics}, \ref{eq:system_delay}), first without considering disturbance (${d(t) \equiv 0}$), and we rely on CBF~(\ref{eq:truck_CBF}).
This system was studied from safety perspective in \cite{HeOrosz2018} without time delay, wherein the following nominal controller was used to execute car following:
\begin{equation}
k_{\rm n}(x) = A (V(D) - v) + B(W(v_{\rm L}) - v).
\label{eq:truck_controller}
\end{equation}
The term with gain $B$ makes the CAT match its speed $v$ to the speed $v_{\rm L}$ of the lead vehicle or the speed limit $v_{\max}$, whichever is smaller, as expressed by the speed policy $W$:
\begin{equation}
W(v_{\rm L}) = \min \{ v_{\rm L}, v_{\max} \}.
\end{equation}
The term with gain $A$ allows the CAT to maintain appropriate distance $D$ by the help of the range policy $V$:
\begin{equation}
V(D) = \min \{ \kappa (D - D_{\rm st}), v_{\max} \},
\end{equation}
meaning that the CAT
shall increase its speed $v$ linearly (with gradient $\kappa$) for distances above a standstill distance $D_{\rm st}$, until reaching the speed limit $v_{\max}$.

Here we endow controller~(\ref{eq:truck_controller}) with safety guarantees by choosing its parameters safely.
When ${\tau = 0}$, we choose ${D_{\rm st} \geq D_{\rm sf}}$ and ${B=\kappa=1/T}$, which makes the controller ${k(t,x) = k_{\rm n}(x)}$ provably safe since it satisfies~(\ref{eq:safety_condition}) with ${\alpha(r) = A r}$.
When ${\tau > 0}$, we maintain safety guarantees by using the predictor feedback controller ${k(t,x,u_t) = k_{\rm n}(x_{\rm p})}$ that satisfies~(\ref{eq:safety_condition_delay}).
First, we demonstrate this controller for the case where the predicted state ${x_{\rm p}=\Psi(\tau,t,x,u_t)}$ is calculated accurately.
This requires information from the lead vehicle about its future intent (namely, ${a_{\rm L}(t+s)}$, ${s \in [0,\tau]}$ that shows up in ${f(t+s,.)}$ in~(\ref{eq:semiflow})), which may be provided by vehicle-to-vehicle (V2V) connectivity.
This intent requirement will be relaxed in the next section.

\begin{table}
\begin{center}
\caption{Parameters for longitudinal control of a connected automated truck.} \label{tab:parameters}
\begin{tabular}{cccc}
Description & Parameter & Value & Unit \\
\hline
input delay  & $\tau$ & 0.5 & s \\
distance gain  & $A$ & 0.4 & 1/s \\
velocity gain  & $B$ & 0.5 & 1/s \\
standstill distance & $D_{\rm st}$ & 5 & m \\
range policy gradient & $\kappa$ & 0.5 & 1/s \\
speed limit & $v_{\max}$ & 20 & m/s \\
safe standstill distance & $D_{\rm sf}$ & 3 & m \\
safe time headway & $T$ & 2 & s \\
\multirow{2}{*}{TISSf parameters}
& $\sigma_0$ & 1 & m/s$^3$ \\
& $\lambda$ & 0.3 & 1/m \\
unmodeled first-order lag  & $\xi$ & 0.25 & s \\
\end{tabular}
\end{center}
\end{table}

Figure~\ref{fig:delay} shows simulation results in an emergency braking scenario where the lead vehicle (red) decelerates harshly until a full stop.
Two controllers are compared for the CAT, with parameters in Table~\ref{tab:parameters}.
When the CAT uses the delay-free control design ${k(t,x) = k_{\rm n}(x)}$ without predictor feedback in the presence of input delay, safety is violated (dashed purple).
As opposed, the predictor feedback controller ${k(t,x,u_t) = k_{\rm n}(x_{\rm p})}$ that accounts for the delay is able keep a safe distance (solid purple), as guaranteed by Theorem~\ref{thm:safety_delay}.
This shows the relevance of addressing the delay carefully, as it may otherwise lead to unsafe behavior.


\section{Robust Safety with Disturbance}

In this section, we provide robust safety guarantees against disturbances via the notion of input-to-state safety.
We also address prediction errors as a part of disturbances.

\subsection{Delay-free Scenario}

Consider control-affine systems with input disturbance:
\begin{equation}
\dot{x}=f(t,x) + g(t,x) \big( u+d(t) \big),
\label{eq:system_disturbance}
\end{equation}
where ${d: \R \to \R^m}$ is bounded: there exists ${\delta \in \R_{\geq 0}}$ such that ${\| d(t) \| \leq \delta}$, ${\forall t \geq 0}$.
The disturbance affects safety as:
\begin{equation}
\dot{h}(t,x,u+d) = \dot{h}(t,x,u) + L_g h(t,x) d,
\label{eq:hdot_disturbance}
\end{equation}
cf.~(\ref{eq:hdot}).
Since $L_g h(t,x)$ characterizes the impact of the disturbance on safety, we use $L_g h(t,x)$ to provide robustness against the input disturbance.
Instead of accounting for the worst-case disturbance as described by \cite{jankovic2018robust} we rely on {\em input-to-state safety} \citep{ames2019issf}.
When a system is input-to-state safe, trajectories are kept within a neighborhood ${S_\delta \supseteq S}$ of the safe set $S$:
\begin{equation}
S_\delta = \{x \in \R^n: h_\delta(x) \geq 0 \},
\label{eq:safeset_disturbance}
\end{equation}
defined with:
\begin{equation}
h_\delta(x) = h(x) + \gamma(h(x),\delta).
\label{eq:hdelta}
\end{equation}
The size of the neighborhood $S_\delta$ depends on the bound $\delta$ of the disturbance, and it is characterized by function ${\gamma: \R \times \R_{\geq 0} \to \R}$ that is continuously differentiable in its first argument and is of class-$\Kinf$ in its second argument (see the expression of $\gamma$ later in~(\ref{eq:gamma})).
Below we outline a method that enables us to make this neighborhood as small as desired and keep safety violations arbitrarily small in the presence of disturbance.
We achieve this by using tunable input-to-state safe control barrier functions (TISSf-CBFs) from \cite{Alan2022}.

\begin{defn} [\cite{Alan2022}]
Function ${h : \R^n \to \mathbb{R}}$ is a {\em tunable input-to-state safe control barrier function (TISSf-CBF)} for~(\ref{eq:system_disturbance}) with a continuously differentiable function ${\sigma: \R \to \R_{>0}}$ if there exists ${\alpha \in \Keinf}$ such that ${\forall t \geq 0}$ and ${\forall x \in \R^n}$:
\begin{equation}
\sup_{u \in \R^m} \dot{h}(t,x,u) > - \alpha(h(x)) + \sigma(h(x)) \|L_g h(t,x)\|^2.
\label{eq:TISSf_CBF_condition}
\end{equation}
\end{defn}
Note the difference from the CBF definition~(\ref{eq:CBF_condition}): an additional term with the Euclidean norm of $L_g h(t,x)$ appears, which provides robustness against disturbance.
By tuning the coefficient $\sigma(h(x))$, one may choose to increase this robustness closer to the safe set boundary or outside the safe set without introducing conservativeness farther inside the safe set.
To achieve this, we choose ${\sigma(h) = \sigma_0 \exp(-\lambda h)}$ with parameters ${\sigma_0, \lambda \in \R_{>0}}$ so that $\sigma(h(x))$ gets larger as $h(x)$ gets smaller.
We remark that a reciprocal coefficient $\epsilon(h)$ is utilized in \cite{Alan2022}, which can be translated to the form given in this study by simply choosing ${\sigma(h)=1/\epsilon(h)}$.
Furthermore, ${\sigma(h) \equiv 0}$ would reduce to the CBF definition~(\ref{eq:CBF_condition}).

The following theorem from \cite{Alan2022} states the robust safety guarantees provided by TISSf-CBFs.
\begin{thm}[\cite{Alan2022}] \label{thm:safety_disturbance}
If $h$ is a TISSf-CBF for (\ref{eq:system_disturbance}) with ${\sigma: \R \to \R_{>0}}$ satisfying ${\sigma'(r) \leq 0}$, ${\forall r \in \R}$ such that ${\alpha^{-1} \in \Keinf}$ is continuously differentiable, then any locally Lipschitz continuous controller ${k : \R \times \R^n \to \R^m}$ satisfying:
\begin{equation}
\dot{h}(t,x,k(t,x)) \geq - \alpha(h(x)) + \sigma(h(x)) \|L_g h(t,x)\|^2,
\label{eq:TISSf_condition}
\end{equation}
${\forall t \geq 0}$ and ${\forall x \in \R^n}$ renders $S_\delta$ with
\begin{equation}
\gamma(h,\delta) = -\alpha^{-1} \bigg( -\frac{\delta^2}{4 \sigma(h)} \bigg)
\label{eq:gamma}
\end{equation}
forward invariant such that ${x(0) \!\in\! S_\delta \!\implies\! x(t) \!\in\! S_\delta}$, ${\forall t \geq 0}$.
\end{thm}

The proof can be generalized from \cite{Alan2022}, and the proof with input delay is given in the next section.
Note that if a controller ${u=k(t,x)}$ satisfies~(\ref{eq:safety_condition}), then robustness to disturbance can be achieved by a simple modification that satisfies~(\ref{eq:TISSf_condition}): ${u=k(t,x)+\sigma(h(x)) L_g h^\top(t,x)}$.


\subsection{Control with Input Delay}

Having the machinery to address input delay and input disturbance separately, now we combine our two approaches to handle both simultaneously.
Consider the control-affine system with input delay and disturbance:
\begin{equation}
\dot{x}(t)=f(t,x(t)) + g(t,x(t)) \big( u(t-\tau)+d(t) \big),
\label{eq:system_disturbance_delay}
\end{equation}
cf.\!~(\ref{eq:system_delay}) and\!~(\ref{eq:system_disturbance}).
We use TISSf-CBFs and predictor feedback to address the disturbance and delay.

The disturbance affects the solution of~(\ref{eq:system_disturbance_delay}) and hence the predicted state ${x_{\rm p}(t) = x(t+\tau)}$.
The predicted state is now given as ${x_{\rm p}(t) = \Psi_d(\tau,t,x,u_t)}$ using the semi-flow ${\Psi_d: [0,\tau] \times \R \times \R^n \times \B \to \R^n}$:
\begin{multline}
\Psi_d(\vartheta,t,x,u_t)
= x + \int_{0}^{\vartheta} \Big(
f \big( t+s, \Psi_d(s,t,x,u_t) \big) \\
+ g \big( t+s, \Psi_d(s,t,x,u_t) \big) \big( u_t(s-\tau) + d(t+s) \big) \Big) {\rm d}s.
\label{eq:semiflow_disturbance}
\end{multline}
We remark that when the disturbance ${d(t+s)}$ or the future dynamics ${f(t+s,.)}$ or ${g(t+s,.)}$ are unknown, ${x_{\rm p}(t) = \Psi_d(\tau,t,x,u_t)}$ is also unknown and needs to be approximated, which inevitably leads to prediction errors.
First, we provide safety guarantees with the theoretical assumption that $x_{\rm p}(t)$ is accurately available, and second, we address prediction errors.
We use the following definition.

\begin{defn}
Function $h : \R^n \to \mathbb{R}$ is a {\em tunable input-to-state safe control barrier function (TISSf-CBF)} for~(\ref{eq:system_disturbance_delay}) with a continuously differentiable function ${\sigma: \R \to \R_{>0}}$ if there exists ${\alpha \in \Keinf}$ such that ${\forall t \geq 0}$, ${\forall x \in \R^n}$ and ${\forall u_t \in \B}$:
\begin{equation}
\sup_{u \in \R^m} \dot{h}(t_{\rm p},x_{\rm p},u) > - \alpha(h(x_{\rm p})) + \sigma(h(x_{\rm p})) \|L_g h(t_{\rm p},x_{\rm p})\|^2,
\label{eq:TISSf_CBF_condition_delay}
\end{equation}
where ${t_{\rm p} = t + \tau}$, ${x_{\rm p}=\Psi_d(\tau,t,x,u_t)}$ with $\Psi_d$ given by~(\ref{eq:semiflow_disturbance}).
\end{defn}

This definition leads to our main result that ensures safety in the presence of both input delay and disturbance.

\begin{thm} \label{thm:safety_disturbance_delay}
If $h$ is a TISSf-CBF for~(\ref{eq:system_disturbance_delay}) with ${\sigma: \R \to \R_{>0}}$ satisfying ${\sigma'(r) \leq 0}$, ${\forall r \in \R}$ such that ${\alpha^{-1} \in \Keinf}$ is continuously differentiable, then any locally Lipschitz continuous controller ${k : \R \times \R^n \times \B \to \R^m}$ satisfying:
\begin{equation}
\dot{h}(t_{\rm p},x_{\rm p},k(t,x,u_t)) \geq - \alpha(h(x_{\rm p})) + \sigma(h(x_{\rm p})) \|L_g h(t_{\rm p},x_{\rm p})\|^2,
\label{eq:TISSf_condition_delay}
\end{equation}
with ${t_{\rm p} = t + \tau}$, ${x_{\rm p}=\Psi_d(\tau,t,x,u_t)}$, ${\forall t \geq 0}$, ${\forall x \in \R^n}$ and ${\forall u_t \in \mathcal{B}}$ renders $S_\delta$ with~(\ref{eq:gamma}) forward invariant such that ${x(t) \in S_\delta}$, ${\forall t \in [0,\tau]}$ ${\implies x(t) \in S_\delta}$, ${\forall t \geq 0}$.
\end{thm}

The proof is given in Appendix~\ref{sec:appdx}.
Inequality~(\ref{eq:TISSf_condition_delay}) can be used to verify input-to-state safety with given controllers or to synthesize robustly safe control laws.


\begin{rem} \label{rem:prediction_error}
Since the semi-flow $\Psi_d$ in~(\ref{eq:semiflow_disturbance}) depends on the disturbance and future dynamics, the future state ${x_{\rm p}(t) = \Psi_d(\tau,t,x(t),u_t)}$ may be unknown.
One may approximate $x_{\rm p}(t)$ by some $\hat{x}_{\rm p}(t)$, for example, as $\hat{x}_{\rm p}(t) = \Psi(\tau,t,x(t),u_t)$ via the semi-flow~(\ref{eq:semiflow}) without disturbance.
Similarly, one may also approximate $t_{\rm p}$ by $\hat{t}_{\rm p}$ (e.g. ${\hat{t}_{\rm p}=t}$) if quantities in~(\ref{eq:TISSf_condition_delay}) are unavailable at the future time $t_{\rm p}$.
These yield prediction errors, and the corresponding controller ${\hat{u} = \hat{k}(t,x,u_t)}$ synthesized via $\hat{x}_{\rm p}$, $\hat{t}_{\rm p}$ will differ from ${u = k(t,x,u_t)}$ synthesized via $x_{\rm p}$, $t_{\rm p}$.
The discrepancy contributes to the disturbance and yields the effective disturbance ${\hat{d}(t) = d(t) + \hat{u}(t-\tau) - u(t-\tau)}$.
However, controllers synthesized via~(\ref{eq:TISSf_condition_delay}) are robust to disturbances by design.
If $\hat{d}(t)$ is bounded and ${\exists \hat{\delta} \in \R_{\geq 0}}$ such that ${\| \hat{d}(t) \| \leq \hat{\delta}}$, ${\forall t \geq 0}$ (e.g. in case of bounded prediction error and Lipschitz continuous controller), then the robust safety guarantees of Theorem~\ref{thm:safety_disturbance_delay} hold with $\hat{\delta}$ instead of $\delta$.
\end{rem}

\begin{rem}
Input-to-state safety allows for safety violations, but they can be made arbitrarily small with large enough $\sigma(h)$ and control effort.
This holds for both Theorems~\ref{thm:safety_disturbance} and~\ref{thm:safety_disturbance_delay} without and with delay.
As such, one may be tempted to use the delay-free control design~(\ref{eq:TISSf_condition}) and treat the entire effect of delay as disturbance, which corresponds to using ${\hat{x}_{\rm p}(t) = x(t)}$ and ${\hat{t}_{\rm p} = t}$.
While this solution exhibits a level of robustness and was successfully implemented in \cite{Alan2022}, the size $\hat{\delta}$ of the disturbance may be large and it may take a large control effort to keep safety violations minimal.
Control effort and safety violations can be significantly decreased by predictors that estimate $x_{\rm p}(t)$ better than ${x(t)}$; see the example below.
\end{rem}

\subsection{Application to Vehicle Control}

\begin{figure}
\begin{center}
\includegraphics[width=8.4cm]{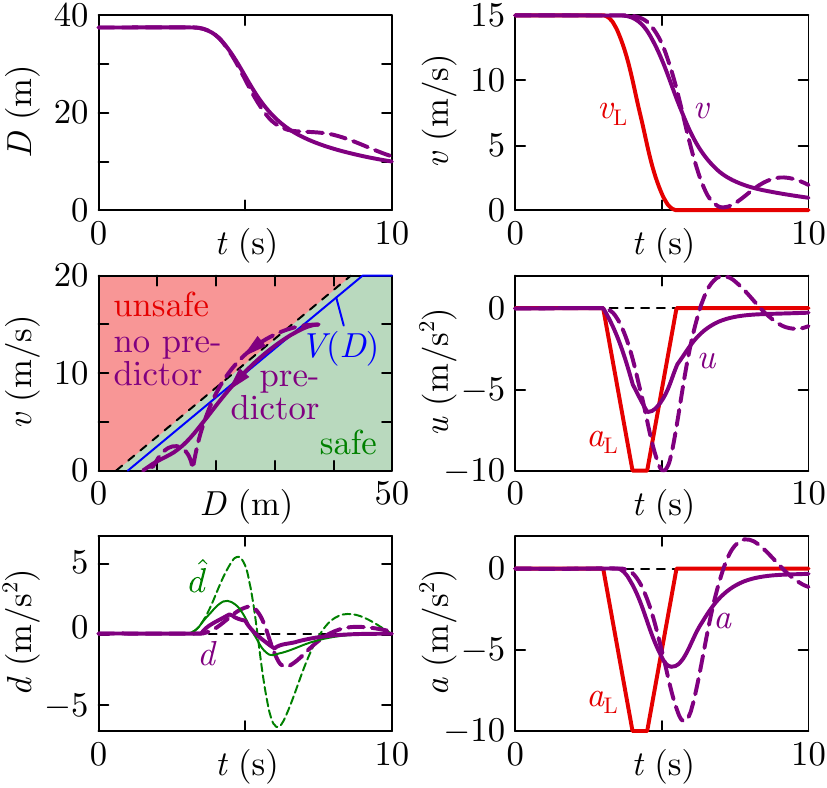}
\vspace{-7pt}
\caption{
Safety-critical control of a CAT with input delay and disturbance.
Two input-to-state safe controllers yield different safety degradation.
While the baseline controller without predictor has larger degradation (dashed), the proposed predictor feedback controller limits it with smaller effective disturbance (solid).}
\label{fig:disturbance}
\end{center}
\end{figure}

Finally, we use the proposed method for controlling the connected automated truck (CAT) described by~(\ref{eq:truck_dynamics}, \ref{eq:system_disturbance_delay}).
We consider a disturbance arising from an additional, unmodeled first-order lag ${\xi \in \R_{> 0}}$:
\begin{align}
\begin{split}   
\dot{a}(t) & = \frac{1}{\xi}(-a(t) + u(t-\tau)), \\
d(t) & = a(t) - u(t-\tau),
\end{split}
\end{align}
where ${a(t) = \dot{v}(t)}$ is the acceleration of the CAT.
Although the first-order lag increases the system dimension, control design still relies on the low-dimensional model~(\ref{eq:system_disturbance_delay}) with a disturbance that will be shown to be bounded in Fig.~\ref{fig:disturbance}.

We consider the disturbance and the lead vehicle's future motion to be unknown (i.e., ${a_{\rm L}(t+s)}$ appearing in ${f(t+s,.)}$ is unavailable for any ${s>0}$).
Thus, we approximate $t_{\rm p}$ by ${\hat{t}_{\rm p} = t}$ and $x_{\rm p}(t)$ by ${\hat{x}_{\rm p}(t) = \hat{\Psi}(\tau,t,x(t),u_t)}$, with a predictor $\hat{\Psi}$ that assumes zero disturbance and constant lead vehicle acceleration (${f(t+s,.) \approx f(t,.)}$):
\begin{multline}
\hat{\Psi}(\vartheta,t,x,u_t)
= x + \int_{0}^{\vartheta} \Big(
f \big( t, \hat{\Psi}(s,t,x,u_t) \big) \\
+ g \big( t, \hat{\Psi}(s,t,x,u_t) \big) u_t(s-\tau) \Big) {\rm d}s,
\end{multline}
cf.~(\ref{eq:semiflow}).
We compare two controllers:
the robust delay-free design ${k(t,x) = k_{\rm n}(x) + \sigma(h(x)) L_g h^\top(t,x)}$  satisfying~(\ref{eq:TISSf_CBF_condition}) and the predictor feedback controller $k(t,x,u_t) = k_{\rm n}(\hat{x}_{\rm p}) + \sigma(h(\hat{x}_{\rm p})) L_g h^\top(\hat{t}_{\rm p},\hat{x}_{\rm p})$ satisfying~(\ref{eq:TISSf_CBF_condition_delay}); both with the nominal control law $k_{\rm n}$ given in~(\ref{eq:truck_controller}).

The resulting closed-loop system is simulated\footnote{See Matlab code at: https://github.com/molnartamasg/safety-critical-control-with-input-delay-and-disturbance.}
in Fig.~\ref{fig:disturbance} for the scenario of Fig.~\ref{fig:delay} with parameters in Table~\ref{tab:parameters}.
The delay-free control design (dashed purple), that treats the effect of the delay as part of the disturbance, yields small safety violations with large control effort.
As opposed, the proposed predictor feedback controller (solid purple) uses significantly smaller control input while keeping the system safe, despite the unmodeled first-order lag and prediction errors.
This is achieved by reducing the effective disturbance $\hat{d}$ relative to the ideal input calculated with ground truth future state $x_{\rm p}$ (thin green lines).
This justifies that even an approximate prediction $\hat{x}_{\rm p}$ may yield better performance than using the current state $x$, and addressing the delay with predictor feedback is beneficial.


\section{Conclusion}

This work addressed the robust safety-critical control of systems with input delay and input disturbance.
Predictor feedback and tunable input-to-state safe control barrier functions were integrated to achieve provable safety guarantees.
The proposed control method was applied to the longitudinal control of a connected automated truck, and showed better performance than delay-free control designs or controllers treating the delay as disturbance.
Future work may involve safety for systems with state delays and robust safety with disturbance observers.




% ...

% ...


% ...

\bibliography{tds2022}

\begin{thebibliography}{19}
\providecommand{\natexlab}[1]{#1}
\providecommand{\url}[1]{\texttt{#1}}
\providecommand{\urlprefix}{URL }
\expandafter\ifx\csname urlstyle\endcsname\relax
  \providecommand{\doi}[1]{doi:\discretionary{}{}{}#1}\else
  \providecommand{\doi}{doi:\discretionary{}{}{}\begingroup
  \urlstyle{rm}\Url}\fi

\bibitem[{Abel et~al.(2019)Abel, Jankovic, and
  Krsti{\'{c}}}]{abel2019constrained}
Abel, I., Jankovic, M., and Krsti{\'{c}}, M. (2019).
\newblock Constrained stabilization of multi-input linear systems with distinct
  input delays.
\newblock \emph{IFAC-PapersOnLine}, 52(2), 82--87.

\bibitem[{Abel et~al.(2020)Abel, Jankovi{\'{c}}, and
  Krsti{\'{c}}}]{abel2020constrained}
Abel, I., Jankovi{\'{c}}, M., and Krsti{\'{c}}, M. (2020).
\newblock Constrained control of input delayed systems with partially
  compensated input delays.
\newblock In \emph{Dynamic Systems and Control Conference}, volume 84270,
  V001T04A006. American Society of Mechanical Engineers.

\bibitem[{Abel et~al.(2021)Abel, Krsti{\'{c}}, and Jankovi{\'{c}}}]{Abel2021}
Abel, I., Krsti{\'{c}}, M., and Jankovi{\'{c}}, M. (2021).
\newblock Safety-critical control of systems with time-varying input delay.
\newblock \emph{IFAC-PapersOnLine}, 54(18), 169--174.

\bibitem[{Alan et~al.(2022)Alan, Taylor, He, Orosz, and Ames}]{Alan2022}
Alan, A., Taylor, A.J., He, C.R., Orosz, G., and Ames, A.D. (2022).
\newblock Safe controller synthesis with tunable input-to-state safe control
  barrier functions.
\newblock \emph{IEEE Control Systems Letters}, 6, 908--913.

\bibitem[{Ames et~al.(2014)Ames, Grizzle, and Tabuada}]{AmesCDC2014}
Ames, A.D., Grizzle, J.W., and Tabuada, P. (2014).
\newblock Control barrier function based quadratic programs with application to
  adaptive cruise control.
\newblock In \emph{53rd IEEE Conference on Decision and Control}, 6271--6278.

\bibitem[{Ames et~al.(2017)Ames, Xu, Grizzle, and Tabuada}]{AmesXuGriTab2017}
Ames, A.D., Xu, X., Grizzle, J.W., and Tabuada, P. (2017).
\newblock Control barrier function based quadratic programs for safety critical
  systems.
\newblock \emph{IEEE Transactions on Automatic Control}, 62(8), 3861--3876.

\bibitem[{Bekiaris-Liberis and Krstic(2013)}]{BekKrs2013}
Bekiaris-Liberis, N. and Krstic, M. (2013).
\newblock \emph{Nonlinear Control Under Nonconstant Delays}.
\newblock SIAM.

\bibitem[{He and Orosz(2018)}]{HeOrosz2018}
He, C.R. and Orosz, G. (2018).
\newblock Safety guaranteed connected cruise control.
\newblock In \emph{21st IEEE International Conference on Intelligent
  Transportation Systems}, 549--554.

\bibitem[{Jankovic(2018{\natexlab{a}})}]{Jankovic2018}
Jankovic, M. (2018{\natexlab{a}}).
\newblock Control barrier functions for constrained control of linear systems
  with input delay.
\newblock In \emph{American Control Conference}, 3316--3321.

\bibitem[{Jankovic(2018{\natexlab{b}})}]{jankovic2018robust}
Jankovic, M. (2018{\natexlab{b}}).
\newblock Robust control barrier functions for constrained stabilization of
  nonlinear systems.
\newblock \emph{Automatica}, 96, 359--367.

\bibitem[{Karafyllis and Krstic(2017)}]{Karafyllis2017}
Karafyllis, I. and Krstic, M. (2017).
\newblock \emph{Predictor feedback for delay systems: {I}mplementations and
  approximations}.
\newblock Birkh{\"{a}}user, Basel.

\bibitem[{{Kolathaya} and {Ames}(2019)}]{ames2019issf}
{Kolathaya}, S. and {Ames}, A.D. (2019).
\newblock Input-to-state safety with control barrier functions.
\newblock \emph{IEEE Control Systems Letters}, 3(1), 108--113.

\bibitem[{Liu et~al.(2020)Liu, Yang, and Ozay}]{necmiye2020}
Liu, Z., Yang, L., and Ozay, N. (2020).
\newblock Scalable computation of controlled invariant sets for discrete-time
  linear systems with input delays.
\newblock In \emph{American Control Conference}, 4722--4728.

\bibitem[{Michiels and Niculescu(2007)}]{michiels2007stability}
Michiels, W. and Niculescu, S.I. (2007).
\newblock \emph{Stability and stabilization of time-delay systems: {An}
  eigenvalue-based approach}.
\newblock SIAM.

\bibitem[{Molnar et~al.(2021)Molnar, Kiss, Ames, and Orosz}]{Molnar2021tcst}
Molnar, T.G., Kiss, A.K., Ames, A.D., and Orosz, G. (2021).
\newblock Safety-critical control with input delay in dynamic environment.
\newblock \emph{arXiv preprint}, (arXiv:2112.08445).

\bibitem[{Nagumo(1942)}]{nagumo1942lage}
Nagumo, M. (1942).
\newblock {\"U}ber die lage der integralkurven gew{\"o}hnlicher
  differentialgleichungen.
\newblock \emph{Proceedings of the Physico-Mathematical Society of Japan}, 24,
  551--559.

\bibitem[{Seiler et~al.(2022)Seiler, Jankovic, and Hellstrom}]{Seiler2022}
Seiler, P., Jankovic, M., and Hellstrom, E. (2022).
\newblock Control barrier functions with unmodeled dynamics using integral
  quadratic constraints.
\newblock \emph{IEEE Control Systems Letters}, 6, 1664--1669.

\bibitem[{Singletary et~al.(2020)Singletary, Chen, and
  Ames}]{singletary2020control}
Singletary, A., Chen, Y., and Ames, A.D. (2020).
\newblock Control barrier functions for sampled-data systems with input delays.
\newblock In \emph{59th IEEE Conference on Decision and Control}, 804--809.

\bibitem[{Vasudevan et~al.(2012)Vasudevan, Shia, Gao, Cervera-Navarro, Bajcsy,
  and Borrelli}]{Vasudevan2012}
Vasudevan, R., Shia, V., Gao, Y., Cervera-Navarro, R., Bajcsy, R., and
  Borrelli, F. (2012).
\newblock Safe semi-autonomous control with enhanced driver modeling.
\newblock In \emph{American Control Conference}, 2896--2903.

\end{thebibliography}

\appendix
\section{Proof of Theorem~\ref{thm:safety_disturbance_delay}}
\label{sec:appdx}

The theorem is conditioned on ${x(t) \in S_\delta}$, ${\forall t \in [0,\tau]}$.
This yields ${x(\tau) \in S_\delta}$ and we prove
${x(\tau) \in S_\delta \implies x(t) \in S_\delta}$, ${\forall t \geq \tau}$.
This is equivalent to
${x_{\rm p}(0) \in S_\delta \implies x_{\rm p}(t) \in S_\delta}$, ${\forall t \geq 0}$.
We begin by shifting argument $t$ in~(\ref{eq:system_disturbance_delay}) to ${t+\tau}$ that yields the evolution of $x_{\rm p}(t)$ in~(\ref{eq:predictedstate}) as:
\begin{equation}
\dot{x}_{\rm p}(t)=f(t_{\rm p},x_{\rm p}(t)) + g(t_{\rm p},x_{\rm p}(t))(u(t)+d_{\rm p}(t)),
\label{eq:system_disturbance_pred}
\end{equation}
with ${d_{\rm p}(t) = d(t+\tau)}$ that satisfies ${\| d_{\rm p}(t) \| \leq \delta}$, ${\forall t \geq 0}$.

Since~(\ref{eq:system_disturbance_pred}) is a delay-free system, the necessary and sufficient condition for ${x_{\rm p}(0) \in S_\delta \implies x_{\rm p}(t) \in S_\delta}$, ${\forall t \geq 0}$ can be given by the theorem of \cite{nagumo1942lage} as:
\begin{equation}
h_\delta(x_{\rm p}) = 0 \implies \dot{h}_\delta(t_{\rm p},x_{\rm p},u+d_{\rm p}) \geq 0.
\label{eq:Nagumo}
\end{equation}
Using~(\ref{eq:hdelta}), we can express $\dot{h}_\delta$ as:
\begin{align}
\dot{h}_\delta(t_{\rm p},x_{\rm p},u+d_{\rm p}) & = \left( 1 + \frac{\partial \gamma}{\partial h}(h(x_{\rm p}),\delta) \right) \dot{h}(t_{\rm p},x_{\rm p},u+d_{\rm p}),
\end{align}
where, according to~(\ref{eq:gamma}), the following holds:
\begin{equation}
\frac{\partial \gamma}{\partial h}(h(x_{\rm p}),\delta)
= -\frac{{\rm d} \alpha^{-1}}{{\rm d} r}\left( -\frac{\delta^2}{4 \sigma(h(x_{\rm p}))} \right) \frac{\delta^2 \sigma'(h(x_{\rm p}))}{4 \sigma(h(x_{\rm p}))^2} > 0,
\end{equation}
since the derivative of ${\alpha^{-1} \in \Keinf}$ above is positive while ${\sigma'(h(x_{\rm p})) \leq 0}$.
Hence,~(\ref{eq:Nagumo}) is equivalent to:
\begin{equation}
h_\delta(x_{\rm p}) = 0 \implies \dot{h}(t_{\rm p},x_{\rm p},u+d_{\rm p}) \geq 0.
\label{eq:Nagumo_modified}
\end{equation}
To prove this, we express $\dot{h}$ as:
\begin{align}
\dot{h}(t_{\rm p},x_{\rm p}&,u+d_{\rm p}) = \dot{h}(t_{\rm p},x_{\rm p},u) + L_g h(t_{\rm p},x_{\rm p}) d_{\rm p},
\label{eq:proof_1} \\
& \geq - \alpha(h(x_{\rm p})) + \sigma(h(x_{\rm p})) \|L_g h(t_{\rm p},x_{\rm p})\|^2
\label{eq:proof_2} \\
& \quad - \|L_g h(t_{\rm p},x_{\rm p})\| \delta \nonumber \\
& \geq - \alpha(h(x_{\rm p})) - \frac{\delta^2}{4 \sigma(h(x_{\rm p}))}
\label{eq:proof_3} \\
& \quad + \left( \sqrt{\sigma(h(x_{\rm p}))} \|L_g h(t_{\rm p},x_{\rm p})\| - \frac{\delta}{2 \sqrt{\sigma(h(x_{\rm p}))}} \right)^2 \nonumber \\
& \geq - \alpha(h(x_{\rm p})) + \alpha(h(x_{\rm p}) - h_\delta(x_{\rm p})),
\label{eq:proof_4}
\end{align}
where we substituted~(\ref{eq:hdot_disturbance}) in~(\ref{eq:proof_1});
we used~(\ref{eq:TISSf_condition_delay}), the Cauchy-Schwartz inequality and ${\| d_{\rm p} \| \leq \delta}$ in~(\ref{eq:proof_2});
we completed the square in~(\ref{eq:proof_3});
and we applied the definition~(\ref{eq:hdelta}, \ref{eq:gamma}) of $h_\delta$ in~(\ref{eq:proof_4}).
From~(\ref{eq:proof_4}), the implication~(\ref{eq:Nagumo_modified}) follows, and the proof is complete.
\hfill $\blacksquare$

\end{document}